\documentclass[a4,11pt]{article}
\usepackage{amsmath}
\usepackage{graphicx}
\usepackage{amssymb} 
\newcommand{\Mat}[1]{{{\boldsymbol{#1}}}}
\newcommand{\abs}[1]{\left\vert#1\right\vert}
\def\be{\begin{equation}}
\def\ee{\end{equation}}
\def\dd{\mathrm{d}}
\title{Equations of motion according to the asymptotic post-Newtonian scheme for general relativity}

\bigskip
\author{
Mayeul Arminjon \\
\small\it Laboratoire ``Sols, Solides, Structures'' (Unit\'e Mixte de Recherche of the CNRS), \\[-1.mm]
\small\it BP 53, F-38041 Grenoble cedex 9, France\\
\small Currently at \it Dipartimento di Fisica, Universit\`a di Bari,
Via Amendola 173, I-70126 Bari, Italy.
}
\date{ }

\begin{document}

\maketitle

\begin{abstract}
We summarize a recent work done on the title's subject. First, we present the asymptotic scheme of post-Newtonian (PN) approximation for general relativity in the harmonic gauge. Then, we discuss the definition of the mass centers and the derivation of equations for their motion, following that scheme. Finally, we briefly analyze the reason why a new term has thus been found in the equations of motion. 

\end{abstract}

\section {Introduction}

The aim of this contribution \footnote
{common text of two talks given at the Third Advanced Research Workshop ``Gravity, Astrophysics and Strings at the Black Sea" (Kiten, Bulgaria, June 13-20, 2005). Submitted to the Proceedings (P. Fiziev and M. Todorov, eds.).
} 
is to summarize a recent work \cite{A36} on the approximate equations of motion relevant to general-relativistic celestial mechanics. The latter equations belong to the heart of the observational test of general relativity (GR), insofar as the first thing which one expects from a theory of gravity is to provide an accurate celestial mechanics. The first part of the paper will be concerned with the construction of the post-Newtonian approximation (PNA). Contrasting the ``standard" scheme of PNA with the ``asymptotic" one, which is used in this work, we will then present the application of the latter scheme to GR in the harmonic gauge and show some of the corresponding PN expansions. In the second part of the paper, the definition of the mass centers will be discussed and the way to derive explicit 1PN equations of motion for the mass centers will be summarized. This derivation uses a recently-developed asymptotic framework for well-separated celestial bodies \cite{A32}. In addition, it is needed to exploit the fact that the main bodies of the solar system are nearly spherical and have approximately a rigid motion. We end the paper by showing the main new term found and discussing the reason why it occurs in this approach and did not in previous works.

\section{Post-Newtonian approximation}
The purpose of the PNA is to get approximate solutions of the Einstein equations for a weakly- gravitating system. In a further step, some integration should provide tractable equations of motion of the mass centers of the different bodies. Since a high accuracy is aimed at, the approximation scheme is an important point. The {\it unknowns} are: the (symmetric) metric tensor $(g_{\mu \nu})$ (ten scalar unknowns), and the independent matter fields: the pressure $p$ plus the coordinate 3-velocity ${\bf u}\equiv \frac{\dd {\bf x}}{\dd t}$, with ${\bf x}=(x^i)$ the spatial position in the coordinate system (thus four scalar unknowns). We consider a barotropic perfect fluid, for which the proper rest-mass density $\rho ^\ast $ and the internal energy density $\Pi $ are functions of the pressure $p$. The {\it equations} are the ten Einstein equations, plus the four equations of the gauge condition---here we adopt the harmonic gauge condition, which is usually used in relativistic celestial mechanics, and which is now indeed officially recommended \cite{Soffel2003} in this domain.

\subsection{Standard PN approximation}\label{StandardPN}
This is the procedure followed by Fock in his classical book \cite{Fock64}. It is essentially equivalent to the one discussed later by Chandrasekhar \cite{Chandra65}. It is still adopted in most of the current textbooks. In this approach, the metric $(g_{\mu \nu})$, as well as the connection and the curvature tensor, are formally expanded in power series of $c^{-1}$, where $c$ is the velocity of light. However, the matter fields  $p, {\bf u} $, etc., are not expanded. As already noted by Futamase \& Schutz \cite{FutaSchutz} and by Rendall \cite {Rendall92}, the latter point leads to a certain difficulty in interpreting the equations obtained. If one searches for asymptotic expansions in the usual sense, then all fields have to be expanded, moreover one has to properly define a small parameter $\epsilon $ and to define a family $(\mathrm{S}_\epsilon)$ of systems, in order that $\epsilon $ can (conceptually) be made arbitrarily small. 

\subsection{Asymptotic scheme of PN approximation} 
  
This was first proposed by Futamase \& Schutz \cite{FutaSchutz}, and this was for GR in the harmonic gauge. They derived the local equations (which are essentially the expansion of Einstein equations), though in a form which is not explicit enough for further use. As a matter of fact, ``global" equations, {\it i.e.} ones for mass centers of extended bodies, were not derived by them, nor seemingly by anyone within this scheme, before the work \cite{A36}. In the present study, we start from Futamase-Schutz' {\it family of initial conditions} (slightly modified):
\begin{equation}\label{PN_IC1}
p^{(\epsilon)}(\mathbf{x})=\epsilon^4 p^{(1)}(\mathbf{x}),
\ee 
\begin{equation}\label{PN_IC2}
\rho^{\ast (\epsilon)}(\mathbf{x})=\epsilon^2 \rho^{\ast (1)}(\mathbf{x}),
\end{equation}
\begin{equation}\label{PN_IC3}
\mathbf{u}^{(\epsilon)}(\mathbf{x})= \epsilon\, \mathbf{u}^{(1)}(\mathbf{x}),
\end {equation}
\be \label{ICmetric-F&S}
\sqrt{- g^{(\epsilon)}}\ g^{(\epsilon)\,ij} ({\bf x}) = \delta _{ij}, \qquad
\left(\sqrt{-g^{(\epsilon)}}\ g^{(\epsilon)\,ij} \right)_{,0} ({\bf x}) =0. 
\ee
[Here, $g \equiv \text{det}(g_{\mu \nu })$, and $(g^{\mu \nu })$ is the inverse matrix of $(g_{\mu \nu })$.] Let $\mathrm{S}_\epsilon$ be the system defined by this initial condition: we have indeed a family ($\mathrm{S}_\epsilon$) of systems.

\subsubsection{Expansion of the matter fields}

Let $[\mathrm{M}]$ and $[\mathrm{T}]$ be the mass and the time units for system S$_1$. We change the mass and the time units for system $\mathrm{S}_\epsilon$, by adopting a large time unit $\ [\mathrm{T}]_\epsilon = [\mathrm{T}]/\epsilon$ and a very small mass unit $[\mathrm{M}]_\epsilon = \epsilon^2 [\mathrm{M}]$, where $[\mathrm{M}]$ and $[\mathrm{T}]$ are the units for system S$_1$ \cite{A23}. After this change, the initial data (\ref{PN_IC1})-(\ref{ICmetric-F&S}) becomes {\it independent of} $\epsilon$, and we have just $ \Mat{\epsilon =c^{-1}}$ (we take $c=1$ in units for system S$_1$). The initial-value problem still depends on $\epsilon$ {\it or more exactly on $\epsilon^2$,} since the Einstein equations involve only the square $c^2=\epsilon^{-2}$. This suggests stating Taylor expansions with respect to $\epsilon^2=c^{-2}$, starting from the {\it zero-order} term:
\be
p' = p'_0 + p'_1 c^{-2}+ O(c^{-4}),\qquad {\bf u}' = {\bf u}'_0 + {\bf u}'_1 c^{-2}+ O(c^{-4}), 
\ee
{\it etc.} In fixed units (the units of system S$_1$), this is rewritten as
\be \label{matter-expansion}
p = [p_0 + p_1 \epsilon^2+ O(\epsilon^4)]\epsilon^4,\qquad {\bf u} = [{\bf u}_0 + {\bf u}_1 \epsilon^2+ O(\epsilon^4)]\epsilon, 
\ee
{\it etc.}
 
\subsubsection{Expansion of the metric}
 
In a weak field, we expect that the metric $\Mat{g} \approx  \Mat{\eta }$, the flat metric. In particular, $ \Mat{g}$ should have an expansion with a first term of order zero (in fixed units):
\be\label{g-order0}
\qquad \Mat{g} =\  _0\Mat{g} + O(\epsilon ^{2}).
\ee
Starting from (\ref{g-order0}), the ``relaxed" Einstein equations (valid in the harmonic gauge) allow us to deduce \cite {A36} that indeed 
\be
\qquad _0\Mat{g} =\  \Mat{\eta }=\mathrm{diag}(-1,1,1,1).
\ee
With the time unit $\ [\mathrm{T}]_\epsilon = [\mathrm{T}]/\epsilon$, this is

\be
\qquad _0\Mat{g}' = \mathrm{diag}(-\epsilon ^{-2},1,1,1) = \mathrm{diag}(-c^2,1,1,1).
\ee
To 1PN order, one postulates Taylor expansions in $c^{-2}$, up to and including the $c^{-2}$ term, {\it e.g.}
\be \label{g00-expansion-c2}
g'_{00 } = -c^2  + \,_1g'_{00 } +\, _2g'_{00 }\,c^{-2}    + O(c^{-4}). 
\ee 
In the fixed starting units, this leads \cite{A36} to the expansion of the metric written by Weinberg \cite{Weinberg}, including the $g_{0i}$'s in $\epsilon ^3$.

\subsubsection{Expanded field equations: general form}

The expansion coefficients: $\,p_0,p_1,\,_1g_{00 }, \,_2g_{00 },\ etc.$, are functions of the position ${\bf x}$ and of the ``dynamical time" $t'\equiv \epsilon t$. They are by definition independent of $\epsilon $. Starting from the expansions of the matter fields and the metric, one deduces expansions for the other quantities, {\it e.g.} for the energy-momentum tensor ${\bf T}$:

\be \label{orders-T}
T^{\mu \nu} = \epsilon ^{n_{\mu \nu }} \left (_0T^{\mu \nu} +\,_1T^{\mu \nu}\,\epsilon ^2 +O(\epsilon ^4)\right ), \quad n_{00}=2, \ n_{i0} = 3, \ n_{ij}=4.
\ee
Again, the coefficients, {\it e.g.} $_0T^{\mu \nu}$ and $_1T^{\mu \nu}$ in (\ref{orders-T}), do not depend on the parameter $\epsilon $. Therefore, if one enters these expansions into the Einstein equations and into the dynamical equations ($T^{\mu \nu }_{;\nu }=0$), one gets each of these equations {\it split to two separate equations,} according to the order in $\epsilon ^2$: this is merely coefficient identification in a polynomial function.\\

The general form of the split equations, {\it i.e.}, with the coefficients $_0T^{\mu \nu},\ _1T^{\mu \nu}$, not with their explicit expressions, is the same as for Weinberg. This is because he too expands ${\bf T}$.

\subsubsection{Explicit expansions for a perfect fluid} 
 
We enter the expansions of the metric and the matter fields, into the expression of ${\bf T}$ for a perfect fluid:
\be \label{T-fluid}
T^{\mu \nu } = (\mu^{\ast} +p) U^\mu U^\nu + p g^{\mu \nu } \quad (c=1)
\ee
($U^\mu\equiv \dd x^\mu/\dd s$ is the 4-velocity,$\quad\mu^{\ast}\equiv\rho^\ast(1+\Pi)$ is the proper energy density). This leads to the explicit expansion of tensor ${\bf T}$. {\it E.g.}:
\be \label{expans-T00}
_0T^{00} = \mu^{\ast}_0, \quad _1T^{00} = \mu^{\ast}_1 +\mu^{\ast}_0({\bf u}_0^2-2\Phi ). 
\ee
($\Phi$ is the Newtonian potential.) Weinberg \cite{Weinberg} does not expand the matter fields $\mu^{\ast}, {\bf u},\  etc.$. He denotes by $\rho $ what we note $\mu^{\ast}$ in Eq. (\ref{T-fluid}), but then he writes, for instance,
\be \label{expans-T00-SW}
_0T^{00} = \rho, \quad _1T^{00} = \rho ({\bf u}^2-2\Phi ).
\ee
One may tentatively interpret the unexpanded matter fields of Eq. (\ref{expans-T00-SW}) within the asymptotic scheme, as being the first-approximation fields. This interpretation means, for instance:
\be
\rho _\mathrm{SW} \equiv \mu^{\ast}_{(1)}\equiv (\mu^{\ast} _0 + \mu^{\ast} _1 \epsilon^2)\epsilon^2,\quad {\bf u}_\mathrm{SW}\equiv {\bf u}_{(1)}\equiv \epsilon ( {\bf u}_0 + {\bf u}_1 \epsilon ^2).
\ee
Then the coefficients $_0T^{00}_\mathrm{SW}$, $_1T^{00}_\mathrm{SW}$ in Eq. (\ref{expans-T00-SW}), do depend on the small parameter $\epsilon$. (This is called a ``composite expansion.") But then, it is difficult to see why the field equations could be split. On the other hand, if one does not split the equations, while expanding some of the independent variables (here the metric is expanded), then one has not enough equations. More precisely, at the 1PN order, the expansion of the metric component $g_{00}$ uses two independent unknowns, Eq. (\ref{g00-expansion-c2}). Hence, for this component, a problem arises \cite{A36} if the corresponding Einstein equation is not split to two equations.

\subsubsection{Explicit expanded equations for a perfect fluid}\label{ExplicitExpansFluid}
 
Inserting the asymptotic expansion of ${\bf T}$, Eq. (\ref{expans-T00}) and the like, into the dynamical equation $T^{\mu \nu }_{;\nu }=0$, gives the equations of fluid dynamics. The zero-order equations turn out to be those of classical fluid dynamics: continuity equation and Euler's equation. {\it They are exact for the zero-order fields!} That continuity equation expresses the mass conservation at the order zero, {\it not} at PN order. However, the order-one time equation, combined with the zero-order equations, allows one to check that mass is conserved also at the 1PN approximation. (In fact, mass is exactly conserved for an isentropic perfect fluid in GR.) Due to the expansion of matter fields, the order-one equations are definitely different from the corresponding equations of Weinberg \cite{Weinberg} , which coincide with those of Chandrasekhar \cite{Chandra65}.

\section{The definition of the mass centers and their equations of motion}
\subsection{Definition of the mass centers: motivation}

In any relativistic gravity, any form of energy must 
\begin{itemize}
	\item contribute to the gravitational field;
	\item be subjected to its action. 
\end{itemize}
Therefore, a question arises: which energy density should one choose as ``the" correct weight function $\phi$ so as to define relevant mass centers?\\

The energy density $\phi$ adopted as the weight function should
\begin{itemize}
	\item i) be well-correlated with the luminous density;
	\item ii) ensure that the velocity of a body's mass center is the average velocity of this body's constituents.
\end{itemize}
Condition ii) is satisfied if and only if $\phi $ satisfies the continuity equation: 
\be
\partial_t \phi +\mathrm{div}(\phi {\bf u})=0.
\ee
Therefore, the rest-mass density $\rho $ (in the global frame), which satisfies these two conditions and which seems to be the only energy density to do so, should be chosen as the weight function $\phi$. A more standard choice [{\it e.g.} Will \cite{Will93}, Eqs.~(6.21) and (6.25)] is to consider a density $\rho '=\rho $ {\it plus:} i) the density of internal energy, ii) the energy density associated with the self Newtonian potential of the relevant body, and iii) the density of the kinetic energy associated with its motion with respect to a local frame attached to the mass center of that body. (However, several well-known works also take ``$\phi \equiv \rho$'', e.g. Fock \cite{Fock64}, Brumberg \cite{Brumberg91}.)  Since $\rho '-\rho $ is of order $\rho\times GM_a/(c^2 r_a)$ where $r_a$ is the size of body $(a)$, the two different choices can lead only to small and, most importantly, nonsecular differences.

\subsection{Formal definition of the mass centers}

The masses and the mass centers are defined with the global rest-mass density $\rho$:
\begin{equation}\label{defmasscent}
  M_a \equiv\int_{\mathrm{D}_a}\rho \ \dd V,\qquad M_a \mathbf{a}\equiv\int_{\mathrm{D}_a}\rho\mathbf{x}\ \dd V,
\end{equation}
where $\mathrm{D}_a(t)$ is the spatial domain occupied by body $(a)$ ($a=1,..., N$) in the considered (harmonic) coordinate system. 

At the 1PN approximation, $\rho$ is approximated by $\rho_{(1)}\equiv \rho _0+ \rho _1 c^{-2}$, hence 
\begin{equation}\label{defPNmass}
  M_a^{(1)}=M^0_a+M_a^1 c^{-2} ,\quad M^0_a\equiv \int_{\mathrm{D}_a}\rho_0
  \ \dd V,\quad M_a^1\equiv\int_{\mathrm{D}_a}\rho_1 \ \dd V,
\end{equation}
\begin{equation}\label{defPNmasscent}
  M_a^{(1)}\mathbf{a}_{(1)}\equiv\int_{\mathrm{D}_a}\rho_{(1)}\mathbf{x}\ \dd V=
  M^0_a\mathbf{a} _{0} +M_a^{1}\mathbf{a} _{1} c^{-2} ,
\end{equation}
with
\begin{equation}\label{defmasscent-ord0-ord1}
  M^0_a\mathbf{a}_{0} \equiv \int_{\mathrm{D}_a}\rho_0\mathbf{x}\ \dd V,\qquad M_a^{1}\mathbf{a}_{1}\equiv \int_{\mathrm{D}_a}\rho_{1}\mathbf{x}\ \dd V.
\end{equation}
Note that $M^0_a$ and $\mathbf{a}_{0} $ are the Newtonian mass and mass center, because the zero-order fields obey the Newtonian equations. The definitions (\ref{defPNmasscent}) and (\ref{defmasscent-ord0-ord1}) seem natural. Also note, however, that the 1PN position of the mass center, $\mathbf{a}_{(1)}$, is not equal to $\mathbf{a} _{0} +\mathbf{a} _{1} c^{-2}$. In other words, $\mathbf{a} _{1}$ is not the 1PN correction to the position of the mass center.

\subsection{1PN equations of motion of the mass centers: general form} 
  
To get the PN equations of motion of the mass centers, the local PN equations of motion (obtained as described in $\S $ \ref{ExplicitExpansFluid}) are integrated in the domain $\mathrm{D}_a$. More precisely, integration of the space components of the local PN equations of motion gives the sought equations for motion of the PN mass center ${\bf a} _{(1)}$ of body $(a)$. At the order zero, we get the Newtonian equation of motion:
\begin{equation}\label{masscent-ord0}
  M^0_a\ddot{{\bf a}}_0=-\int_{\mathrm{D}_a} \rho \nabla \Phi^{(a)} dV 
\end{equation}
($\Phi^{(a)}$ is the external part, for body $(a)$, of the Newtonian potential $\Phi $). For the 1PN correction, we get:
\be\label{delta-addot}
\ddot{{\bf a}}_{(1)} - \ddot{{\bf a}}_0 = \frac{-\dot{{\bf I}}^{a}+{\bf J}^{a}+{\bf K}^{a}-M_a^1 \ddot{{\bf a}}_0}{c^2M_a^0},
\ee
where ${\bf I}^{a},{\bf J}^{a},{\bf K}^{a}$ are integrals of PN fields over ${\mathrm{D}_a}$, hence are {\it structure-dependent}. Thus, {\it a priori}, the equations of motion are so.

\subsection{Good separation between celestial bodies}
For many self-gravitating systems, in particular for our solar system, the main bodies are well separated. This means that one may introduce a small parameter thus:
\begin{equation}\label{eta}
  \eta_0\equiv\max_{a \neq b}\frac{r_b}{\abs{\mathbf{a-b}}} \ll 1
  \qquad (r_b\equiv \mathrm{radius\ of\ body\ domain\ D}_b).
\end{equation}
To account for this, we assume that the 1PN equations are accurate enough for the system of interest, S. In other words, S is replaced by the corresponding 1PN system $\mathrm{S}'$. (This is in order to avoid worrying with {\it two} small parameters, $\epsilon$ and $\eta $.) Then, we build a family of 1PN systems, (S$'^\eta$), with $\mathrm{S}'^{\eta _0}=\mathrm{S}'$. This is defined by a family of initial conditions which ensure that
	\begin{itemize}

	\item $(r_{ab}^0)^\eta \equiv \abs{\mathbf{a}_0^{\eta}-\mathbf{b}_0^{\eta}}=
  	\mathrm{ord}(\eta^{-1}) \quad \mathrm{for\ }a \neq b$,

	\item $  (\dot{{\bf a}}_0)^\eta = \mathrm{ord}(\eta^{1/2})$,

	\item $(\Mat{\Omega}^{(a)})^\eta = \mathrm{ord}(\eta^{1/2}), \ \Mat{\Omega}^{(a)}\equiv \mathrm{rotation\ velocity\ }.$\\
(The zero-order velocity field is assumed to be a rigid rotation, see just below.)

 	\end{itemize}

\subsection{Justification of assuming rigidly-rotating bodies}

Even at Newtonian order, rigid rotation cannot be exact, due to the time-varying influence of the other bodies, which produces {\it tides}. However, if one writes Euler's equation for a rotating body in a well-separated system, one finds that, neglecting $O(\eta^3)$, the body's internal equilibrium is not affected by the other bodies. This accuracy for the internal equilibrium turns out to be enough to get equations of motion up to $\eta ^3$ {\it included}. To this accuracy, it is hence consistent to assume rigidly-rotating bodies.

\subsection{Quasi-spherical bodies}

In addition to having nearly rigid rotation, the big celestial bodies are nearly spherical, but not exactly so: cf. the case of Jupiter. This can be described in an asymptotic framework by coupling the good separation with the quasi-sphericity: we assume that
\be \label{quasi-spheric}
\abs{\gamma ^{(a)}_i - \gamma ^{(a)}_k} = O(\eta^2) \quad (a = 1,..., N;\ i,k=1,2,3),
\ee
where the $\gamma ^{(a)}_i$ 's are the eigenvalues of the inertia tensor of body $(a)$.\\

As a result, it is shown that the rotation velocity {\it rate} verifies
\be \label{Omega-dot=O(eta^3)}
\dot{\Mat{\Omega}}^{(a)} = O(\eta^3), 
\ee 
which may be neglected for the $\ \mathrm{ord}(\eta ^3)$ equations of motion.

\subsection{Equations of motion (well-separated, rigidly-rotating, quasi-spherical bodies)} 
 
In this framework, integrals ${\bf I}^{a},{\bf J}^{a},{\bf K}^{a}$, entering the equation for 1PN correction to the acceleration of the mass center, Eq. (\ref{delta-addot}), can be computed up to $\eta ^3$ included. Inserting the result in this equation gives the explicit equations of motion in the asymptotic scheme. The Newtonian and 1PN equations are thus {\it separated.} To compare with the standard Lorentz-Droste \cite{Lorentz-Droste} (or Einstein-Infeld-Hoffmann \cite{EIH}) equations, one needs to group both together. This turns out to be feasible. As a result, we find \cite{A36} that the acceleration of the mass center of the generic body $(a)$ is exactly the Lorentz-Droste acceleration ${\bf A}_a^\text{LD}$, {\it plus} a 1PN correction ${\bf A}_a^\mathrm{ns}$ that cancels if all bodies are exactly spherically symmetric (this term is likely to be extremely small in the solar system), {\it plus} a 1PN correction depending on the rotation velocity of the body and on its internal structure:
\begin{eqnarray}\label{deltaA}
\ddot{{\bf a}}_{(1)} - {\bf A}_a^\text{LD} & = & \frac{\gamma _a}{c^2M^0_a} \left[ 6(\omega^{(a)})^2 + {\bf \Omega}^{(a)}.{\bf \Omega}^{(a)} \right].\ddot{{\bf a}}_0+ {\bf A}_a^\mathrm{ns}+O(\eta ^4) + O(c^{-4}).
\end{eqnarray}
Here, $\gamma _a$ is the spherical inertia moment, which does depend on the density profile, hence on the internal structure; and $(\omega^{(a)})^2 \equiv \Omega^{(a)}_{jk} \Omega^{(a)}_{jk}/2$ is the square of the angular velocity. [${\bf \Omega}^{(a)}.{\bf \Omega}^{(a)}$ is the spatial tensor (in Cartesian coordinates) with $ik$ component $\Omega^{(a)}_{ij} \Omega^{(a)}_{jk}$. Note that $\ddot{{\bf a}}_0$ is the Newtonian acceleration (\ref{masscent-ord0}) of the body considered.] We note that this new term is order 3 in $\eta$. It is clearly a self-acceleration term. This new term does not seem to be negligible for the giant planets, and, the author conjectures, it will lead to secular effects.

\subsection{Why is the new term here and was absent in previous works?}

There are essentially two different approaches to the equations of motion.\\

1. There are works based on the local integration of the 1PN field equations obtained in the global reference frame for a perfect-fluid system, as is based the present work. The previous works that belonged to this category ({\it e.g.} Fock \cite{Fock64}, Misner {\it et al.} \cite{MTW}, Spyrou \cite{Spyrou78}, Will \cite{Will93}, Brumberg \cite{Brumberg91}):
\begin{itemize}
 
\item used the standard (Fock-Chandrasekhar) approximation scheme summarized in Subsection \ref{StandardPN}. The present author believes to have shown that this scheme is not really compatible with an asymptotic interpretation \cite{A36,O1}. Of course, both the standard and the asymptotic PNA are just approximations, and thus, in a sense, they are both ``wrong''. However, it seems to this author that the asymptotic scheme has serious chances to approximate accurately the exact behaviour---because, precisely, it is fully consistent with asymptotic analysis;

\item did not involve either an asymptotic description of the good separation between bodies. (This point is discussed in Ref. \cite{A32}.) The same lack led, in the first version \cite{A25-26} of the equations of motion of the scalar theory investigated by the author, to neglect terms which were later found \cite{A32} to be numerically significant.\\

\end{itemize}

2. More recently, there have been works based on multipole series expansions with local frames (Damour, Soffel, and Xu \cite{DSX91,DSX92}, Racine and Flanagan \cite{RacineFlanagan}). These works have led to tractable equations of motion only for two models: the ``monopole model", which gives the Lorentz-Droste acceleration, and the ``monopole-dipole model", which adds a (negligible) spin-dependent term. Here we note that these authors use a rather complex kind of multipole series expansion, so that it is difficult to get an asymptotic estimate of the omitted remainder. \\
\bigskip

{\bf Acknowledgement.} I am grateful to the organizers of this Series of Workshops, in particular to P. Fiziev, R. Rashkov, and M. Todorov, for the nice and fruitful atmosphere that they manage to instigate in Kiten. It is also a pleasure to thank F. Selleri and F. Romano for their hospitality in Bari.

\end{document}